\documentclass[jkps,twocolumn,fleqn,showpacs,showkeys]{revtex4}     

\pdfoutput=1
\usepackage{amsmath,amsfonts}
\usepackage{epsfig,subfigure}
\usepackage{ifpdf}
\usepackage{hyperref}
\usepackage{amsmath}
\usepackage{graphicx}
\usepackage{color}
\usepackage{natbib}
\usepackage{multirow}
\usepackage{kotex}
\usepackage{times}

\newcommand{\msun}{M_\odot}
\newcommand{\be}{\begin{equation}}
\newcommand{\ee}{\end{equation}}
\newcommand{\bea}{\begin{eqnarray}}
\newcommand{\eea}{\end{eqnarray}}

\newcommand{\etal}{{\it et al.}}

\begin{document}
\setcounter{page}{1}

\title{Gravitational Wave Searches for Aligned-Spin Binary Neutron Stars Using Nonspinning Templates}

\author{Hee-Suk \surname{Cho}}
\email{chohs1439@pusan.ac.kr}
\affiliation{Department of Physics, Pusan National University, Busan 46241, Korea}

\author{Chang-Hwan \surname{Lee}}
\affiliation{Department of Physics, Pusan National University, Busan 46241, Korea}

\date{\today}

\begin{abstract}
We study gravitational wave searches for merging binary neutron stars (NSs).
We use nonspinning template waveforms towards the 
signals emitted from aligned-spin NS-NS binaries, in which the spins of the NSs are aligned with 
the orbital angular momentum.
We use the TaylorF2 waveform model, which can generate inspiral waveforms emitted from aligned-spin compact binaries.
We employ the single effective spin parameter $\chi_{\rm eff}$ to represent the 
effect of two component spins ($\chi_1, \chi_2$) on the wave function.
For a target system, we choose a binary consisting of the same component masses of $1.4 \msun$
and  consider the spins up to $\chi_i= 0.4$,
We investigate fitting factors of the nonspinning templates 
to evaluate their efficiency in gravitational wave searches for the aligned-spin NS-NS binaries. 
We find that the templates can achieve the fitting factors exceeding $0.97$ only for the signals in the range of $-0.2 \lesssim \chi_{\rm eff} \lesssim 0$.
Therefore, we demonstrate the necessity of using aligned-spin templates not to lose the signals outside that range.
We also show how much the recovered total mass can be biased from the true value depending on the spin of the signal.

\end{abstract}



%=======	Title		================================	

\maketitle

%=======	Intro		================================	
\section{Introduction}
Recently, three gravitational wave (GW) signals have been
detected by the two Advanced LIGO detectors \cite{GW1,GW2,GW3}.
According to the observing dates, those signals were 
named as GW150914, GW151226, and GW170104.
Detailed analyses showed that 
they were emitted from
merging binary black holes \cite{GW1PE1,GW1PE2,GW12PE}.
On the other hand, 
although GW signals from black hole-neutron star (NS) binaries or NS-NS binaries 
have not yet been detected \cite{GWBNS}, those signals are also
expected to be captured in the near future by the advanced detector network \cite{ALIGO,AVirgo,KAGRA}
(for an overview of NS-NS merger rates, see \cite{Aba10,GWBNS}).
They will tell us more information about the nature of NSs,
such as equation of states and mass distribution  \cite{Lat07,Pra13,Lee11,Lee14}.

In this work, we investigate fitting factors of nonspinning waveform templates
for GW signals from aligned-spin NS-NS binaries based on  the Advanced LIGO detectors. 
The fitting factor represents
the best-match of template waveforms to a signal waveform \cite{Apo95}. 
It is generally used to evaluate the
efficiency of templates in GW searches. 
For example, if ${\rm FF}\simeq 0.97$, that means a loss of detection rates of $\sim 10\%$,
because the detection rate is proportional to the cube root of the signal-to-noise ratio ($\rho$), $\rho^{1/3}$,
and $\rho \propto {\rm FF}$.
The studies on the efficiency of nonspinning templates for aligned-spin signals have been carried out in the past works \cite{Pri14,Dal14,Cap16,Cho16b}, but those works have only considered binary black holes or black hole-NS binaries.
Several authors have also performed such a study towards 
the precessing-spin signals \cite{Bro12,Aji14,Har14,Dal15}.
Here, we focus on aligned-spin NS-NS binaries with the spins of $-0.4 \leq \chi_i \leq 0.4$ 
and  calculate fitting factors and biases using nonspinning templates.
In particular, we give a detailed interpretation for the physical property of the results.

For merging NS-NS binaries, the GW frequencies are most likely to get out of the sensitivity frequency band of Advanced LIGO
before the systems reach the ultra-relativistic regime.
Therefore, the post-Newtonian approximation is valid for modeling the
gravitational waveforms  emitted from those binaries \cite{Bla14}.
Here, we adopt a simple post-Newtonian waveform model, TaylorF2,
in which the wave function is defined in the Fourier domain.

%=======	Section 2: method	================================

%=================   data analysis	====================

\section{GW search analysis}

%=================   section 2-1 : FF and bias	====================
In signal processing, the matched filtering can be the most efficient method for 
signals of known shape buried in stationary Gaussian noise.
Since the inspiral waveforms emitted from merging  NS-NS binaries 
can be modeled almost accurately by the post-Newtonian approximation,
the matched filter can be used in the GW search analysis.
The match between a detector data stream $x(t)$, consisting of a GW signal and stationary Gaussian noise,
and a waveform $h(t)$ is defined as 
\be \label{eq.match}
\langle x | h \rangle = 4 {\rm Re} \int_{f_{\rm low}}^{\infty}  \frac{\tilde{x}(f)\tilde{h}^*(f)}{S_n(f)} df,
\ee
where the tilde denotes the Fourier transform of the time-domain waveform, $S_n(f)$ is the power spectral density (PSD) of the detector noise, and the low frequency cutoff ($f_{\rm low}$) can be determined by the PSD curve. 
In this work, we use the zero-detuned, high-power noise PSD of Advanced LIGO \cite{apsd}
and assume $f_{\rm low}=10$ Hz.
If the normalized waveform is defined as $\hat{h} \equiv h/ \langle h| h\rangle^{1/2}$,
the signal-to-noise ratio can be obtained by $\rho=\langle x | \hat{h} \rangle$.

In order to describe aligned-spin binary systems in circular orbits, 
the wave function should incorporate five extrinsic parameters (luminosity distance of the
binary, two angles defining the sky position of the binary
with respect to the detector, orbital inclination, and
wave polarization), two arbitrary constants (coalescence
time $t_c$ and coalescence phase $\phi_c$), and four intrinsic
parameters (two masses and two spins).
On the other hand, the extrinsic parameters only scale the wave
amplitude, hence do not affect our analysis \cite{Cho14}.
Therefore, we do not consider those extrinsic parameters in this work.

We define the overlap ($P$) by the match between a normalized signal $\hat{h}_s$ and a normalized template $\hat{h}_t$ \cite{Cho13}:
\be\label{eq.overlap}
P =  \max_{t_c,\phi_c}\langle \hat{h}_s | \hat{h}_t \rangle.
\ee
Here, the maximization over  $t_c$ and $\phi_c$ can be
easily performed by using certain analytic techniques \cite{All12}.
Since we consider only nonspinning waveforms as templates, 
the overlaps can be distributed in the two-dimensional mass parameter space as
\be\label{eq.2d overlap}
P(m_1, m_2) =  \max_{t_c,\phi_c}\langle \hat{h}_s | \hat{h}_t(m_1, m_2) \rangle.
\ee
The fitting factor (FF) is defined as the best-match between $\hat{h}_s$ and a set of $\hat{h}_t$ \cite{Apo95}:
\be
{\rm FF}=\max_{\lambda}\langle \hat{h}_s | \hat{h}_t(\lambda) \rangle,
\ee
where $\lambda$ represents the physical parameters considered in the template space.
The connection of the fitting factor to the overlap can be given as
\be\label{eq.FF}
{\rm FF} = \max_{\lambda}P(\lambda).
\ee

If we use a complete template waveform model that can produce exactly the same shape as the signal waveform,
we obtain the fitting factor equal to 1, and 
the parameter values recovered by the templates are the same as the true values.
However, typically, a waveform model cannot be complete;
thus, the recovered values  are likely to be biased from the true values systematically,
and the fitting factor should be lower than 1.
Therefore, in our overlap distributions, the recovered masses can be biased 
because we use nonspinning template waveforms towards the spinning signals.
The bias can be easily determined by the distance from the true value ($\lambda_0$) to the recovered value ($\lambda^{\rm rec}$):
\be\label{eq.bias}
b =   \lambda^{\rm rec} - \lambda_0.
\ee
The bias corresponds to a
systematic error in the GW parameter estimation.
As the efficiency of a template waveform model for the search is evaluated by the fitting factor,
its validity for the parameter estimation can be  examined by the bias \cite{Cut07,Fav14,Cho15d}.

We determine the fitting factor and the bias in the following way \cite{Cho15c,Cho15d,Cho16b}.
First, we repeat a grid search around $\lambda_0$ 
until we find the crude location of $\lambda^{\rm rec}$ in the overlap surface. 
Note that, we use the chirp mass $M_c \equiv
(m_1 m_2)^{3/5}/M^{1/5}$ and the symmetric mass ratio
$\eta \equiv m_1m_2/M^2$, where $M=m_1+m_2$, because those parameters are much more efficient
than the component masses in our analysis \cite{Cho15a}.
Next, we estimate the size of the contour $\bar{P} \equiv P/ P_{\rm max} = 0.995$,
where $P_{\rm max}$ is the maximum value in the overlap surface.
Finally, we find (almost) the exact location of $\lambda^{\rm rec}$ by performing a $51 \times 51$ grid search
in the region of $\bar{P} > 0.995$ and choose the overlap value at that location as  the fitting factor.

%%%%%%%%%%%%%%%%%%%% waveform   %%%%%%%%%%%%%%%%%%%%%

%\section{waveform model for aligned-spin binaries}

The waveform function of TaylorF2 is given as
\be \label{eq.TaylorF2}
h(f)=Af^{-7/6} e^{i \Psi(f)},
\ee
where $A$ is the wave amplitude that consists of the masses and the extrinsic parameters.
Since we use the normalized waveforms, the amplitude does not affect our analysis.
The wave phase is expressed as
\begin{equation} \label{eq.Psi}
\Psi(f)=2\pi ft_c -2 \phi_c -{\pi\over 4} + {3 \over 128 \eta v^5} \phi(f),
\end{equation}
where $\phi(f)$ is given by the post-Newtonian expansion  with the expansion parameter $v= [\pi f M]^{1/3}$.
The coefficients of the phase equation are expressed as functions of $\eta$ and the two component spins 
$\chi_1, \chi_2$ with $\chi_i \equiv S_i/m_i^2$, $S_i$ being the spin angular momentum of the $i{\rm th}$ compact object. 
We consider the post-Newtonian expansion up to 3.5 pN order  where the spin terms
are included up to 2.5 pN order \cite{Aru09}.
%%%%%%%%%%%%%%%%%%%% waveform   %%%%%%%%%%%%%%%%%%%%%

\section{result}
For a target system, we choose a NS-NS binary consisting of the same component masses of $1.4\msun$.
We consider the spins up to $\chi_i = 0.4$,
which corresponds to the spin of the fastest-spinning millisecond pulsar observed so far \cite{Hes06}.
Note that, the spins of known pulsars in double neutron star
systems are below 0.04 \cite{Kra09}.
In order to produce our template waveforms, we vary only the mass parameters fixing the spins to be $\chi_1=\chi_2=0$.
We then calculate two-dimensional overlap distributions by using Eq.~(\ref{eq.2d overlap})
and obtain fitting factors and biases by using Eqs.~(\ref{eq.FF}) and (\ref{eq.bias}), respectively.

\begin{figure}[t]
\includegraphics[width=8.0cm]{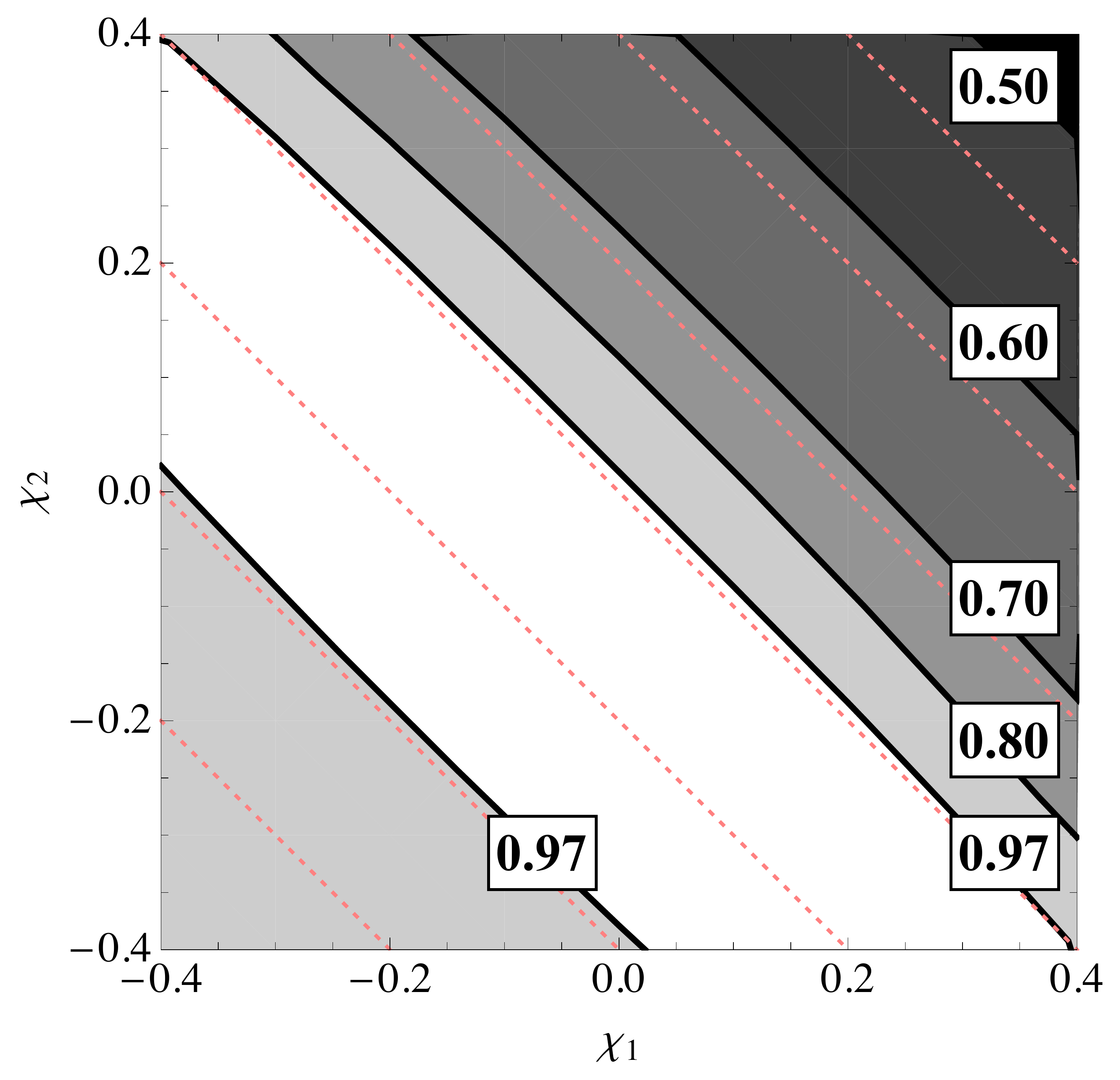}
\caption{Fitting factors calculated by using nonspinning templates for aligned-spin NS-NS signals with the same component masses of $1.4\msun$. The dotted lines indicate the lines of constant effective spin, $\chi_{\rm eff}=\{-0.3, -0.2, ..., 0.3\}$ from bottom left.}\label{fig.ff}
\end{figure}

In Fig.~\ref{fig.ff}, we show the fitting factors of nonspinning templates (cf. Fig~6 of \cite{Aji11a}). 
One can easily find that the fitting factor contours are symmetric about the line of $\chi_1=\chi_2$,
and they are coincide with the lines of constant total spin.
Generally, in aligned-spin systems, the effect of the two component spins  
can be represented by a single effective parameter \cite{Poi95}.
Several definitions for the single parameter have been introduced by Refs. \cite{Dam01,Aji11a,Aji11,San10,Kha16} 
We adopt the simplest one defined by $\chi_{\rm{eff}} = (m_1 \chi_1+m_2 \chi_2)/M$,
which has been introduced to model the phenomenological template families \cite{San10,Aji11}
and used in the recent studies similar to this work \cite{Cho16b,Cho17}.
For equal mass systems, 
the lines of constant total spin are coincide with the lines of constant effective spin.
Therefore, the pattern of the contours shows that the fitting factors mainly depend on the effective spin rather than the component spins.

On the other hand, the fitting factors show a significant discrepancy between the positive and the negative effective spins.
Typically, a positively (negatively) aligned-spin signal tends to be recovered by a nonspinning template having a larger (smaller) value of $\eta$
than the true value of $\eta$, i.e., $\eta^{\rm rec} > \eta_0   \ (\eta^{\rm rec} < \eta_0)$, and the size of the bias ($b_{\eta}$) depends on the spin of the signal.
On the other hand, the parameter value of $\eta$ is physically restricted to the range of $0 \leq \eta \leq 0.25$
(Note that the unphysical value of $\eta$ implies complex-valued masses),
and this restriction results in a rapid fall-off of the fitting factor when $\eta^{\rm rec}$ reaches the physical boundaries (for more details, see \cite{Dal14, Cho16b}).
In the region of $\chi_{\rm eff} < 0$, as the magnitude of the effective spin increases,
$\eta^{\rm rec}$ decreases but does not reach the boundary $0$ in our spin range 
because this boundary is sufficiently far from the true value ($\eta_0=0.25$).
Therefore, the fitting factor gradually decreases with increasing $|\chi_{\rm eff}|$.
However, in the region of $\chi_{\rm eff} > 0$,
$\eta^{\rm rec}$ is always equal to $0.25$, i.e., $b_{\eta}=0$ because $\eta_0 \leq \eta^{\rm rec} \leq 0.25$,
and consequently the fitting factors have much lower values. 
In conclusion, the nonspinning templates are efficient for GW searches, i.e., ${\rm FF}\geq 0.97$,
for equal mass NS-NS binaries only with the spins of $-0.2 \lesssim \chi_{\rm eff} \lesssim 0$.
Note that, if we choose a binary having sufficiently asymmetric masses so that
$\eta^{\rm rec}$ does not reach the boundaries $0$ and $0.25$ in the regions of $\chi_{\rm eff} < 0$ and $\chi_{\rm eff} > 0$, respectively,
then the fitting factors gradually decrease in both the regions (e.g., see Fig.~2 of \cite{Cho16b}).

\begin{figure}[t]
\includegraphics[width=8.0cm]{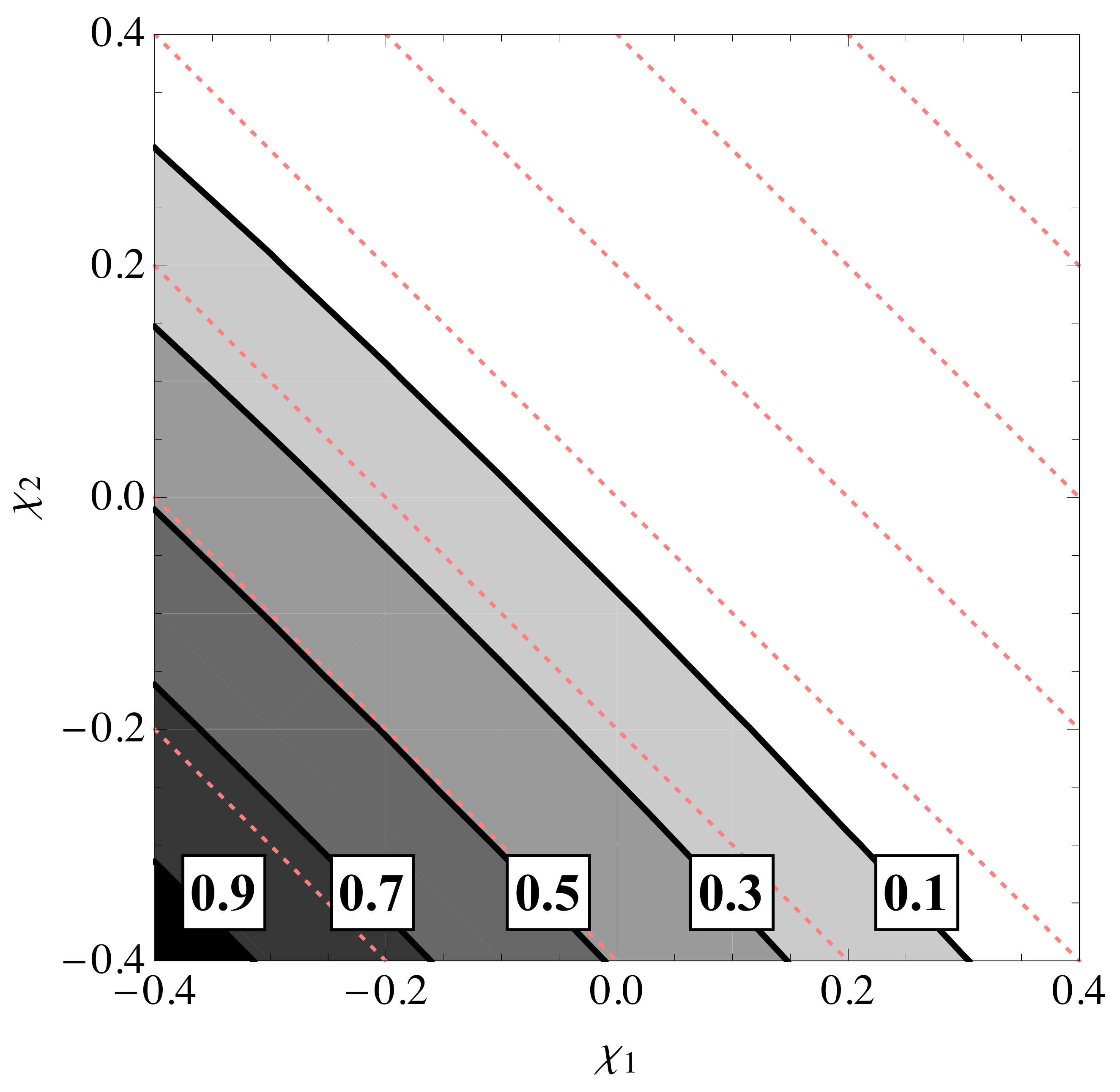}
\caption{Biases for the total mass ($b_M/\msun$) induced by using nonspinning templates for aligned-spin NS-NS signals with the same component masses of $1.4\msun$. The dotted lines are the same as in Fig. \ref{fig.ff}.}\label{fig.bias}
\end{figure}

One can infer how the recovered total mass can be biased by understanding the orbital motion of spinning binary systems.
When the spin is negatively aligned with the orbital angular
momentum, the spin-orbit coupling makes the binary's
phase evolution slightly faster, hence hastens the onset of
the plunge phase, as compared to its nonspinning counterpart \cite{Cam06}. 
For that reason, the negative effective spin decreases
the length of the waveform, as compared to the nonspinning
case, and such a waveform best matches the one
produced by a higher mass nonspinning binary.
In a positively aligned-spin system, the spin-orbit coupling makes exactly the opposite effect.
Therefore, a positive (negative) aligned-spin binary
can be recovered by a lower (higher) mass nonspinning
template. 
In Fig.~\ref{fig.bias}, we show the biases for the total mass of the system, $b_M$.
As described above, one can find the positive biases in the region of $\chi_{\rm eff} < 0$.
However, there are almost no biases in the region of $\chi_{\rm eff} > 0$.
This is due to the constraint on the $\eta$ space.
Typically, as $\chi_{\rm eff}$ increases from 0, the position of $\{M^{\rm rec}, \eta^{\rm rec}\}$ 
tends to move to the upper left hand side from the position of $\{M_0, \eta_0\}$ in the $M-\eta$ plane.  
However, for equal mass binaries, 
since $\eta^{\rm rec}$ is always unbiased due to the physical boundary in this region,
its companion $M^{\rm rec}$ is also unbiased
although the $M$ space is unrestricted here.

%%%%%%%%%%%%%%%%%%%%%%%%%%
\section{summary and discussion}
We studied the efficiency of nonspinning templates in
GW searches for aligned-spin NS-NS binaries.
We assumed that the target binary system had the same component masses of $1.4 \msun$
and calculated fitting factors and biases $b_M$ of nonspinning templates for aligned-spin signals 
with the spins in the range of $-0.4 \leq \chi_i \leq 0.4$.
We found  that both the fitting factor and the bias strongly depend on the
effective spin ($\chi_{\rm eff}$) rather than the component spins,
and we confirmed that unequal mass binaries also show a similar behavior.
In particular, we found a significant discrepancy  
between the positive and the negative effective spins.
In the region of $\chi_{\rm eff}<0$, as the magnitude of the effective spin increases, 
the fitting factor decreases gradually, and the bias ($b_M$) increases steadily up to $+1 \msun$ at $\chi_1=\chi_2=-0.4$.
On the contrary, in the region of $\chi_{\rm eff}>0$,
the fitting factors have much lower values due to the effect of the physical boundary of $\eta$,
and almost no biases are shown for all the signals in that region.

We demonstrated that nonspinning templates can achieve the fitting factors exceeding $0.97$
for aligned-spin NS-NS binaries in the range of $-0.2 \lesssim \chi_{\rm eff} \lesssim 0$;
thus, they are efficient for GW searches only for the signals in that range.
In order not to lose GW signals outside that range,
one should take into account the spin parameters in the template waveforms.
The detection pipeline of Advanced LIGO for NS-NS binaries has been using aligned-spin templates,
where the magnitude of the component object's spin is limited to $\chi_i \leq 0.05$ \cite{GWBNS}.

%=======	Acknowledgements ===========================	
%

\section*{ACKNOWLEDGMENTS}
HSC was supported by the National Research Foundation of Korea (NRF) grant funded by the
Korea government (MSIP) ( No. 2016R1C1B2010064 and No. 2015R1A2A2A01004238).
CHL was supported by the National Research Foundation of Korea (NRF) grant funded by the
Korea government (MSIP) (No. 2015R1A2A2A01004238 and No. 2016R1A5A1013277).
This work used the computing resources at the KISTI Global Science Experimental Data Hub Center (GSDC).
%

%
%
%=======	Bibliography     ===========================	
%

\end{document}